\documentstyle[12pt]{article}
\def\singlespace {\smallskipamount=3.75pt plus1pt minus1pt
                  \medskipamount=7.5pt plus2pt minus2pt
                  \bigskipamount=15pt plus4pt minus4pt
                  \normalbaselineskip=15pt plus0pt minus0pt
                  \normallineskip=1pt
                  \normallineskiplimit=0pt
                  \jot=3.75pt
                  {\def\smallskip {\vskip\smallskipamount}}
                  {\def\medskip   {\vskip\medskipamount}}
                  {\def\bigskip   {\vskip\bigskipamount}}
                  {\setbox\strutbox=\hbox{\vrule
                    height10.5pt depth4.5pt width 0pt}}
                  \parskip 7.5pt
                  \normalbaselines}
\def\middlespace {\smallskipamount=5.825pt plus1.5pt minus1.5pt
                  \medskipamount=11.25pt plus3pt minus3pt
                  \bigskipamount=22.5pt plus6pt minus6pt
                  \normalbaselineskip=22.5pt plus0pt minus0pt
                  \normallineskip=1pt
                  \normallineskiplimit=0pt
                  \jot=5.825pt
                  {\def\smallskip {\vskip\smallskipamount}}
                  {\def\medskip   {\vskip\medskipamount}}
                  {\def\bigskip   {\vskip\bigskipamount}}
                  {\setbox\strutbox=\hbox{\vrule
                    height15.75pt depth6.75pt width 0pt}}
                  \parskip 7.25pt
                  \normalbaselines}
\def\dblspc {\smallskipamount=7.5pt plus2pt minus2pt
                  \medskipamount=15pt plus4pt minus4pt
                  \bigskipamount=30pt plus8pt minus8pt
                  \normalbaselineskip=30pt plus0pt minus0pt
                  \normallineskip=2pt
                  \normallineskiplimit=0pt
                  \jot=7.5pt
                  {\def\smallskip {\vskip\smallskipamount}}
                  {\def\medskip   {\vskip\medskipamount}}
                  {\def\bigskip   {\vskip\bigskipamount}}
                  {\setbox\strutbox=\hbox{\vrule
                    height21.0pt depth9.0pt width 0pt}}
                  \parskip 15.0pt
                  \normalbaselines}

\def\eps{\epsilon}

\def\be{\begin{equation}}

\def\j-{\J_-}

\def\ee{\end{equation}}

\def\bearr{\begin{eqnarray}}
\def\bearrs{\begin{eqnarray*}}
\def\eearr{\end{eqnarray}}
\def\eearrs{\end{eqnarray*}}
\def\barr{\begin{array}}
\def\earr{\end{array}}

\def\non\non{\nonumber}
\def\nn8{\nonumber\\[15pt]}
\def\l{\left}
\def\r{\right}

\begin{document}
\begin{center}
{\Large \bf Behaviour of the Centrifugal Force\\[6pt]
and of Ellipticity for a Slowly Rotating Fluid\\[6pt]
Configuration with Different Equations of State
}\\[25pt]
{\bf Anshu Gupta, Sai Iyer and A. R. Prasanna\\[2pt]
Physical Research Laboratory\\[2pt]
Ahmedabad 380 009, India}\\[30pt]
{\bf \underline{Abstract}}\\[10pt]
\end{center}
\middlespace
We have evaluated the centrifugal force acting on a fluid
element and the ellipticity of the fluid configuration, which is
slowly rotating, using the Hartle-Thorne solution for different equations of
state.  The centrifugal force shows a maximum in every case,
whereas the reversal in sign could be seen in only one case, and
the system becomes unstable in other cases. The
ellipticity as calculated from the usual definition shows
maxima, whereas the definition obtained from the equilibration of the
inertial forces, shows a negative behaviour, indicating that the
system is prolate and not oblate.  This prolate shape of the
configuration is similar to the one earlier found by Pfister and
Braun for a rotating shell of matter, using the correct centrifugal
force expression for the interior.  The location of the
centrifugal maxima gets farther away from the Schwarzschild
radius $\l( R_s \r)$ as the equation of state gets softer.  \\

{\bf Introduction} 

\noindent One of the most important aspects in the discussion of the
dynamics of a rotating neutron star is the equation of state of matter
contained within. There have been a large number of studies in this
regard starting from the pioneering paper of Oppenheimer and Volkoff
\cite{opvol} 
in 1939 and later by Pandharipande (1971) \cite{pand1} and several others over
the last twenty five years. In spite of all the efforts put in, it is
still enigmatic as to the correct nature of the equation of state and 
its transitions within, as it involves the actual short range
interactions about which there exists no exact theory as yet. However,
there have been several different equations of state, considering the
nuclear interactions at various levels and potentials, and it is
important to see the effects that they may have in the actual structure
and dynamics of a neutron star.\\

\noindent It is indeed well known that a rotating body breaks spherical
symmetry and the equilibrium configuration would be a Maclaurin spheroid
or a Jacobi ellipsoid due to ellipticity arising from the anisotropy of
the equilibrating forces. Our interest is to consider the dynamics of
the interior of ultra compact bodies in the realm of general relativity
as the gravitational potential is quite large. Though the original
formulation of general relativity does not express the equilibrium in
terms of forces, a recent 3+1 formulation allows one to introduce the
language of forces without affecting any of the relativistic effects. 
On the other hand, through the language of forces one gets some new 
results which were not apparent earlier. One of the interesting new
features is the realisation that the centrifugal force acting on a fluid
element of the configuration has a maximum value and that it also
reverses its direction at a value of $r$, outside the event horizon 
{\cite{abp}, \cite{asp}}. Further, it has also been seen that the 
ellipticity of the
spheroidal configuration attains a maximum \cite{asp}, a result which was
earlier shown by Miller et al.
\cite{Miller}. As the change in ellipticity would have to do with 
the shape of the
compact object, it is indeed interesting to check how the 
ellipticity and centrifugal force change with different
equations of state. As shown earlier in the case of homogeneous
distribution, the centrifugal force reverses for a configuration with
$R/R_s \approx 1.45$, and the ellipticity for the same configuration
is maximum at $R/R_s \approx 2.75$.\\

\noindent In this, we extend this study to the case of
an inhomogeneous density distribution with four different equations of
state as given below:\\

\begin{itemize}
\item[(A)] Pandharipande (hyperonic matter) \cite{pand2},
\item[(B)] Wiringa, Fiks and Fabrocini's beta-stable model: UV14+UVII
\cite{wir},
\item[(C)] Walecka's model corresponding to pure neutron matter
\cite{wal}, 
and 
\item[(D)] Sahu, Basu and Datta's model based on the chiral sigma model
\cite{sbd}.
\end{itemize}

\noindent Model (A) is Pandharipande's (hyperonic matter) EOS which 
studies the
behaviour of dense matter using a many body theory based upon the
variational approach. Hyperons are considered one of the baryonic
constituents of the neutron star's interior. Model (B) considers the
main constituents of neutron star matter to be neutrons,
protons, electrons and muons. This is one of the three models given 
by Wiringa, Fiks and Fabrocini which includes
three-nucleon interactions using a non-relativistic approach based 
on the variational method.
Walecka's model (C), which
characterizes the effective interaction by the meson parameters 
(masses and the coupling constants), is for
pure neutron matter and emphasises the scalar and vector meson
exchange interactions by considering the relativistic approach.
Model (D) is a field theoretical EOS for neutron rich matter in beta
equilibrium based on the chiral sigma model.\\

\noindent Model A is the softest equation of state among the considered
EOS, whereas the most stiff EOS is model D. B and C are intermediate
EOS of which C is the stiffer of the two.\\

\noindent Using the ACL formalism of 3+1 conformal slicing of the space
time
\begin{equation}
ds^2 = dl^2 + g_{0 0} (dt + 2 \omega_{\alpha} dx^{\alpha})^2,
\end{equation}
\noindent with $dl^2$ representing the positive definite metric of the
absolute 3-space ${\tilde g}_{\mu \nu} dx^{\mu} dx^{\nu}$ (where Greek
indices take values from 1 to 3 and ${\tilde {}}$ denotes the quantities
in absolute 3-space), the four force acting on a fluid element
of a slowly rotating perfect fluid distribution, represented by
the spacetime metric
\begin{equation}
ds^2 = -e^{2 \nu} dt^2 + e^{2 \psi} d{\hat \phi}^2 + e^{2 \mu_1} d r^2 +
e^{2 \mu_2} d \theta^2 \;,
\end{equation}
in a locally non rotating frame (LNRF), is given by \cite{asp}
\begin{equation}
f_0 = \Phi^{-1} (\rho + p) {\tilde U}^\mu \partial_\mu U_0 + h^\mu_0
p_{, \mu}\\[8pt]
\end{equation}
\begin{equation}
f_\alpha  = \Phi^{-1} (\rho + p) \left [ {\tilde U}^\mu {\tilde \nabla}_
\mu {\tilde U}_\alpha + \frac{M_0^2}{2 \Phi} \partial_\alpha \Phi \right
]
+ h^\mu_\alpha p_{, \mu}
\end{equation}

Here $\Phi$ represents the gravitational potential $\l( -g_{oo}
\r)$, $p$ the pressure, $\rho$ the energy density and $U^\mu$
the velocity four vector of a fluid element.  An overhead tilde
represents the quantity in the projected 3-space of the optical
reference geometry. 
The terms ${\tilde U}^\mu {\tilde \nabla}_\mu {\tilde
U}_\alpha$ and $(M_0^2/2\Phi) \partial_\alpha \Phi$ represent the
centrifugal $(F_{cf})$ and gravitational acceleration $(F_{gf})$
respectively and have the form 
\begin{equation}
\begin{array}{lll}
F_{cf} &=& e^{2 \psi + 2 \nu} (\Omega - \omega)^2 (\psi' - \nu') \left [
e^{2 \nu} - e^{2 \psi} (\Omega - \omega)^2 \right ]^{-1}\\[8pt]
F_g &=& e^{2 \nu} \nu',
\end{array}
\end{equation}
for metric (2).\\

\noindent The Newtonian force balance equation 

\begin{equation}
g_{equator} - a \Omega^2 = g_{pole} (1 - e^2)^\frac{1}{2}
\end{equation}
\noindent may be expressed in general as
\begin{equation}
F_{ge} - F_{cf} = F_{gp} (1 - e^2)^{1/2},
\end{equation}

\noindent with `$F_{ge}$' and `$F_{gp}$' representing the accelerations 
at equator and pole respectively.

\noindent In the limit of slow rotation, the ellipticity is given by
\begin{equation}
\epsilon = \frac{1}{2} e^2,
\end{equation}
where $e$ is the eccentricity of the spheroid. Using equations (7) and 
(8), ellipticity is seen to be

\begin{equation}
\epsilon = \frac{1}{2} \left(1 - \left[\frac{F_{cf} - F_{ge}}{F_{gp}}
\right]^2\right).
\end{equation}

\noindent For the Hartle-Thorne \cite{harthor} metric
\begin{eqnarray}
F_{cf} &=& r^2 {\bar\omega}^2 (1/r - \nu_0'/2),\\[8pt]
F_{ge} &=& \frac{1}{2} e^{\nu_0} [\nu_0'(1+2h_0-h_2)+2h_0'-h_2']
,\\[8pt]
F_{gp} &=& \frac{1}{2} e^{\nu_0} [\nu_0'(1+2h_0+2h_2)+2h_0'+2 h_2']
,
\end{eqnarray}
\noindent yielding for the ellipticity function:
\begin{equation}
\epsilon = 3(h_2 + h_2'/\nu_0') + \frac{r^2 {\bar\omega}^2}{e^{\nu_0}}
\left(2/{r \nu_0'} - 1\right).
\end{equation}

{\large Results and Discussions :}\\

Table 1 shows the location of extrema for the centrifugal force 
as well as for the ellipticity and their respective values for
both the Hartle-Thorne definition $\l( \bar{\eps}_{H-T} \r)$ and our
definition $\l( \bar{\eps} \r)$.  Figures (1)-(3) give the
plots of centrifugal force $\bar{F}_{cf}$ and the two
ellipticities (expressed in dimensionless units $J^2/M^4$ and
$J^2/M^5$ respectively) as a function of $R/R_s$, $R_s$ being
the Schwarzschild radius.\\

As is seen, the centrifugal force keeps increasing as the
configuration size gets smaller and then attains a maximum
somewhere between $R/R_s = 2.1$ and $2.3$, for different
equations of state.  However, unlike in the case of a homogeneous
spheroid where the centrifugal force reverses sign at $R = 1.45
R_s$, with these different equations of state the reversal is
seen only in the case of Wiringa et al. model (B), at $R \simeq
1.454\; R_s$.  In other cases the equilibrium configuration
becomes unstable before reaching the value $R = 1.5 R_s$, which
in fact is the radius of the orbit of the particle for which the
centrifugal force is zero in the Schwarzschild space time.\\

Considering the nature of ellipticity it appears that the
behaviour is different for the inhomogeneous distribution than
the case of homogeneous distribution.  As seen from Fig.~2,
the ellipticity $\l( \bar{\eps} \r)$ keep reducing as the
configuration gets smaller, becomes zero (meaning that the shape
becomes spherical) and further on gets to a negative value.
However, the negative value attains a minimum and then again the
ellipticity starts increasing.  \\

The fact that the ellipticity starts decreasing and further
becomes negative, could be due to the reason that the force
balance equation used in defining the ellipticity is perhaps
true only for a homogeneous distribution of the fluid, whereas
we have in the above varying density configurations.  However, a
point that needs to be checked carefully is that when Pfister
and Braun \cite{pfister} used the correct centrifugal force expression for
obtaining the solution for the interior of a mass shell, they
found that a proper boundary fit of the exterior and the interior
solutions for the shell, was possible only if the configuration
is prolate rather than oblate.  Here, in our approach also we
start from the equilibration of the forces within the framework
of general relativity and get prolate configuration for
distributions with inhomogeneous density.\\

It is also worth noting that for the same configurations, when
ellipticity is defined in terms of the radii of the object with
constant surface density embedded in a 3-dimensional flat space,
a la Hartle-Thorne\cite{harthor}, one gets the ellipticity
maximum as was in the case of a purely homogeneous configuration.
In this case, the location of the maxima changes with the
equation of state, shifting inwards as the equation of state gets
stiffer. As the equation of state of any configuration describes
the pressure-density relation, the equilibrium of the
configuration in a sense depicts the balancing of the various
forces like gravity, material binding and the centrifugal force.
 As the equations of state gets stiffer the intra-nucleonic
forces, which are effectively repulsive at very short range
become larger, thus requiring the configuration to get more
compact before similar behaviour of ellipticity extrema is
attained.  However, a matter of concern is the result concerning
the difference in behaviour of the ellipticity function in the
two different treatments, for the inhomogeneous distribution
while the conventional Hartle-Thorne way of defining it via
embedding in a 3-flat geometry shows the function to be
positive, defining it through the balancing of inertial forces a
la Maclaurin and Newton, shows it to be negative.  It is
important to look deeper into this question to ascertain whether
the treatment of defining inertial forces for a fluid
configuration has to be different from the approach used for a
single particle dynamics, particularly for inhomogeneous
distributions. \\

\newpage
\begin{center}
{\bf Figure Captions}\\
\end{center}
\begin{itemize}
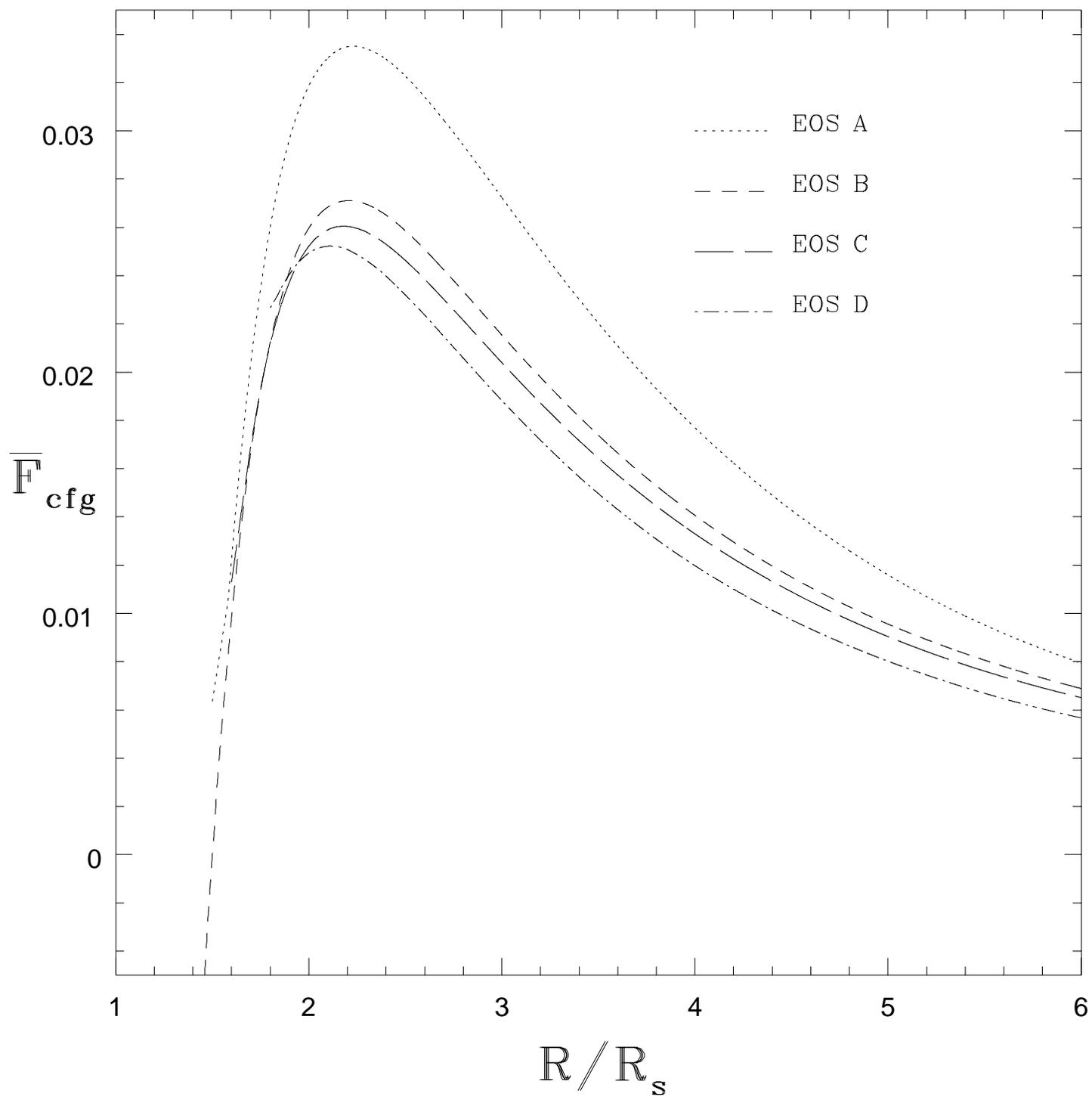
\item[Fig. 1] Centrifugal force ${\bar F}_{cf}$ in units of 
$(J^2/ M^5)$ for decreasing values of radius $R$ in terms 
of Schwarzschild radius $R_s$ for various equations of state. 
EOS A is the softest and EOS D represents the stiffest among 
the four considered equations of state.
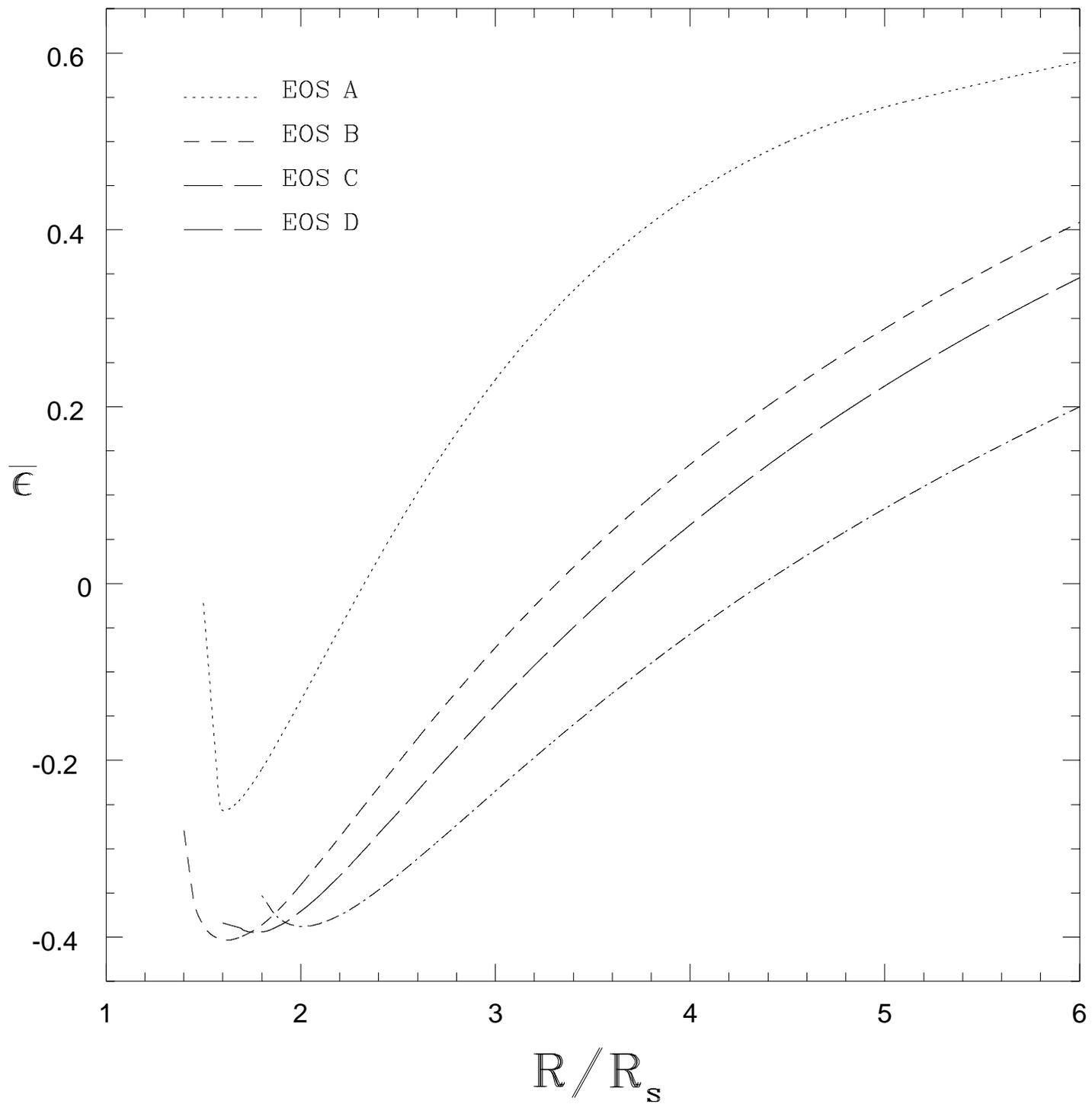
\item[Fig. 2] Ellipticity ${\bar\epsilon}$
derived in optical reference geometry, in units of $J^2/M^4$,
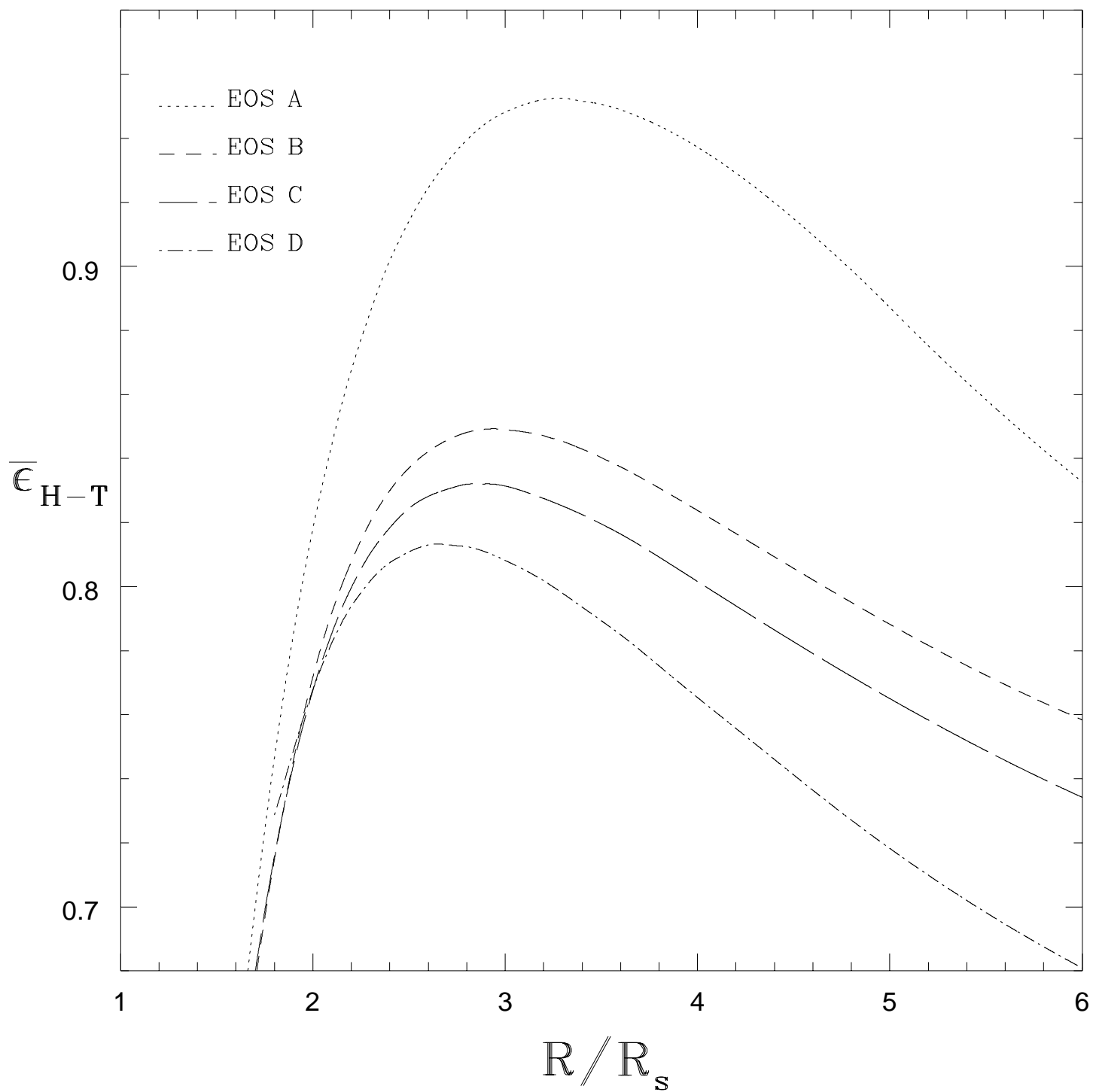
\item[Fig. 3] Ellipticity ${\bar\epsilon}_{H-T}$ as defined by 
Hartle and Thorne, in units of $J^2/M^4$.
\end{itemize}
\newpage
\begin{center}
{\bf Table Captions}\\
\end{center}
\begin{itemize}
\item[Table 1] Location of extrema for the centrifugal force 
($\bar F_{cf}$) and ellipticity ($\bar \epsilon$) (column 7, 5) and 
their values (column 6, 4) for the equations of state (Models A, B, C, D)
considered in this paper as well as for the homogeneous distribution.
Column 8 gives the location of reversal in $\bar \epsilon$.
The radius is expressed in terms of Schwarzschild radius $R_s (= 2 M)$,
whereas the ellipticity and the centrifugal force are expressed in the 
dimensionless units of $J^2/M^5$ and $J^2/M^4$, respectively.
\end{itemize}
\newpage
\def\bfl{\begin{flushleft}}     \def\efl{\end{flushleft}}
\def\bfr{\begin{flushright}}    \def\efr{\end{flushright}}
\def\bc{\begin{center}}         \def\ec{\end{center}}
\begin{center}
\bf Table 1\\ [1cm]
\begin{tabular}{|l |l |l| l|l |l |l| l|}  \hline \hline 
EOS&$\bar{\epsilon}_{H-T}$ & $R_{\bar{\epsilon}_{H-T}}$ & 
$\bar{\epsilon}$ & $R_{\bar{\epsilon}}$ & $\bar{F}_{cf}$
& $R_{\bar{F}_{cf}}$& $R_{\bar{\epsilon}_{rev.}}$ 
\\[4mm] \hline \hline
A&0.953&3.278&-0.257&1.603&0.0335&2.234&2.297\\[4mm]
B&0.849&2.948&-0.403&1.610&0.0271&2.204&3.301\\[4mm]
C&0.832&2.857&-0.395&1.769&0.0261&2.180&3.631\\[4mm]
D&0.813&2.625&-0.388&2.014&0.0252&2.112&4.343\\[4mm]
Homo.&0.761&2.3&1.207&2.75&0.0157&2.1&\\ \hline \hline
\end{tabular}
\end{center}
\end{document}